\documentclass[twocolumn]{elsart5p}

\usepackage{graphicx}
\usepackage{amssymb}
\newcommand{\be}{\begin{eqnarray}}
\newcommand{\ee}{\end{eqnarray}}
\begin{document}
\begin{frontmatter}

\title{Gamma-ray absorption in the microquasar SS433}

\author{M. M. Reynoso\thanksref{label1}\thanksref{lab1}}
\address{Departamento de F\'{\i}sica,
Facultad de Ciencias Exactas y Naturales\\
Universidad Nacional de Mar del Plata, Funes 3350, (7600) Mar del
Plata, Argentina}
\thanks[label1]{\textsl{E-mail address}: mreynoso@mdp.edu.ar (M.M. Reynoso).}
\thanks[lab1]{Fellow of CONICET, Argentina}

\author{H. R. Christiansen}
        \address{State Univesity of Cear\'a, Physics Dept., Av. Paranjana 1700, 60740-000
Fortaleza - CE, Brazil}
 \author{G. E. Romero\thanksref{label2}}
 \address{Instituto Argentino de Radioastronom\'{\i}a, CONICET, C.C.5,
(1894) Villa Elisa, Buenos Aires, Argentina and Facultad de Ciencias
Astron\'omicas y Geof\'{\i}sicas, Universidad Nacional
 de La Plata, Paseo del Bosque, 1900 La Plata, Argentina}\thanks[label2]{Member of CONICET, Argentina}


\begin{abstract}
%

We discuss the gamma-ray absorption in the inner region of the
microquasar SS433. Our investigation includes several contributions
to the opacity of this system. They result from the ambient fields
generated by the primary star, possibly an A-type supergiant, and a
very extended disk around the black hole. Besides the sharp and
dramatic absorption effect that occurs every time the star crosses
the emission zone, we find in the UV photon field from the extended
disk an important source of absorption for very high energy
gamma-rays. This results in periodic gamma-ray observational
signatures.

%
\end{abstract}
\begin{keyword}
gamma-rays: theory \sep X-rays: binaries \sep radiation mechanisms:
non-thermal \sep stars: winds, outflows
\end{keyword}

\end{frontmatter}


\section{INTRODUCTION}
SS433 is a particularly interesting X-ray binary system that
consists of a donor star feeding mass to a black hole
\cite{Fabrika100} in orbit with a period of $13$ d. From the
vicinity of the compact object two oppositely directed jets are
launched developing regular precession with a period of $162$ d.
These jets can be considered as `dark' \cite{gallo} because their
main power output is given by their kinetic luminosity, $L_{\rm
k}\sim 10^{39}$ erg s$^{-1}$. Indeed, the ejected matter  has
determined the deformation of the nebula W50 that surrounds the
SS433 system \cite{dubner}.

Gamma-ray emission from similar objects has been confirmed very
recently \cite{LSI, aharonian, CYG}, so the study of absorption in
the complex case of SS433 is important to characterize the possible
high energy emission that this source might present to instruments
such as GLAST and the new Cherenkov arrays MAGIC II and VERITAS. For
a recent discussion of the absorption in the case of other systems
like LS I +61 303 and LS 5039, see for instance, Ref. \cite{hugo,
romero2007, boettcher}.

Regardless of any specific gamma-ray emission process that could
operate in SS433, in the present work, we shall focus on the
possible sources of gamma-ray absorption taking place mainly in the
inner region and in the immediate vicinity of the binary. There, the
interactions with photons and matter can be important and cause
significant absorption. In particular, if the emission is originated
near the compact object, we find a dramatic absorption effect
occurring every time the companion star eclipses the emission zone.
Furthermore, at very high energies ($E_\gamma\gtrsim 50$ GeV), we
predict a very important absorption effect caused mainly by the UV
emission from the extended disk around the accretion disk, and
secondly by the mid-IR emission from the same disk.

We have based our work on the set of parameters that are currently
believed to describe the system SS433 after more that 20 years of
observation and study. These parameters are summarized in the next
section. The resulting absorption signatures are presented in Sect.
3, and a discussion is left for Sect. 4, where we analyze the
possible detection of the specific absorption features.

\section{THE SOURCE}

\begin{figure}[h]
\includegraphics[trim = 24mm 55mm 26mm 67mm, clip, width=8cm,angle=0]{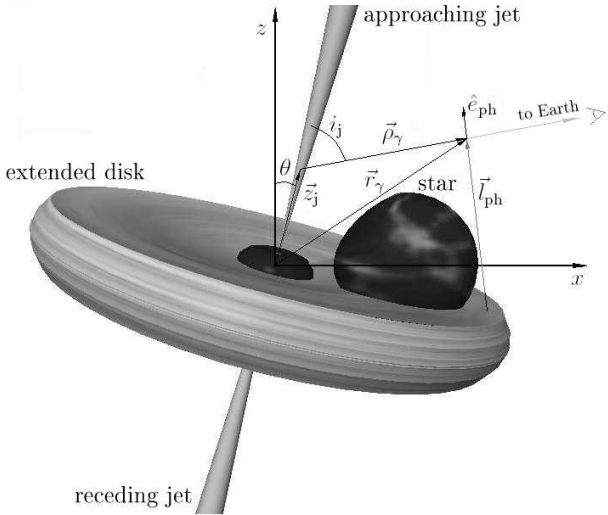}
  \vspace{-10mm}
\caption{Schematic view of SS433. The \textit{approaching} jet is
most of the time closer to the line of sight and the
\textit{receding} one is oppositely directed.} \label{Figrender}
\end{figure}

The mass loss rate in the jets of SS433 is $\dot{m}_{\rm j}= 5\times
10^{-7}{ M}_\odot {\rm yr}^{-1}$ and their bulk velocity is $v_{\rm
b}\approx 0.26$c. The normal to the orbital plane makes an angle
$\theta\approx 21^\circ$ with the approaching jet and an angle
$i=78^\circ$ with the line of sight. The line of sight then makes a
angle $i_{\rm j}$ with the jet which is time-dependent because of
precession (see Fig.\ref{Figrender}).


A thick expanding disk wind believed to be fed by the supercritical
accretion encloses also the star \cite{Zwitter}. According to Ref.
\cite{Fabrika100}, the disk {\textbf wind} has a half opening angle
$\alpha_{\rm w} \approx 30^\circ $, a mass loss rate $\dot{M}_{\rm
w}\approx 10^{-4}{ M}_\odot{\rm yr}^{-1}$, and a velocity $v_{\rm
w}\sim 1500 {\rm \ km \ s}^{-1}$. Emission of UV photons from this
extended disk will be considered using a blackbody distribution with
$T_{\rm UV}=21000$ K following Ref. \cite{Gies02}, while mid-IR
emission will be taken into account using a fit for the flux density
presented in Ref. \cite{Fuchs05}.

The spectral identification of the primary star has been difficult
because of the presence of the extended disk, since the star is
often partially or totally obscured by it. After convenient
observations at specific configurations of precessional and orbital
phases it has become quite clear that the star is an A-supergiant
\cite{Chere,Hillwig,Barnes}. We assume the mass of the components as
derived from INTEGRAL observations \cite{Chere}, $M_{\rm bh}= 9 {
M}_\odot$ and $M_\star= 30 { M}_\odot$ for the black hole and the
star respectively, which correspond to an orbital separation $a
\simeq 79 \ { R}_\odot$ for a zero-eccentricity orbit as is the case
for SS433. Since the star is believed to fill its Roche lobe, the
implied radius according to Ref. \cite{Eggleton} is $R_\star=0.49
q^{-2/3}a\left[0.6 \ q^{-2/3}+ \ln(1+q^{-1/3})\right]^{-1}\simeq 38
{ R}_\odot$, where $q=M_{\rm bh}/M_\star$. We shall consider the
emission of soft photons from the star through an associated
blackbody distribution with an effective temperature $T_{\star}=
8500$ K \cite{Hillwig,Barnes}.

\section{OPACITY TO GAMMA-RAY PROPAGATION}

Attenuation of the putative gamma radiation emitted by this source
would be mainly due to $e^-e^+$ pair production arising from
$\gamma\gamma$ annihilation, particularly with lower energy photons
coming from the extended disk and from the star. Matter, on the
other hand, also offers a significant target for gamma-rays emitted
by the system. We shall calculate also the effect of $\gamma N$
interactions around the star and the extended disk.

When a gamma-ray of energy $E_\gamma$ travels a  distance
$d\rho_\gamma$ in a photon field, there is a differential optical
depth associated with it. Absorption correspondingly arises from the
gamma-ray interaction with soft photons of energy $E$. Assuming that
the former runs following $\hat{e}_{\gamma}$ and the latter are
directed along $\hat{e}_{\rm ph}= (\sin{\theta'}\cos
\phi',\sin{\theta'}\sin \phi',\cos{\theta'})$, the corresponding
optical depth is given by \cite{GS67}, \be d\tau_{\gamma\gamma}= (1-
\hat{e}_\gamma \cdot \hat{e}_{\rm ph}) n_{\rm ph}
\sigma_{\gamma\gamma} \ d\rho_\gamma \ dE  \ d \cos\theta' \ d\phi'.
\label{dtau}
 \ee
Here, $n_{\rm ph}$ is the density of the soft photons per solid
angle and soft photon energy units. The angles $\theta'$ and $\phi'$
are taken in a convenient coordinate system and the cross section
for the process $\gamma\gamma \rightarrow e^+e^-$ is given by
 \be
\sigma_{\gamma\gamma}(E_\gamma,E)=\frac{\pi r_0^2}{2}(1-\xi^2)\times \nonumber\\
\left[2\xi(\xi^2-2)+
(3-\xi^4)\ln\left(\frac{1+\xi}{1-\xi}\right)\right] \label{sigee},
 \ee
 where $r_0$ is the classical electron radius and
 \be
\xi=\left[1-\frac{2(m_e c^2)^2}{E_\gamma E(1-\hat{e}_\gamma \cdot
\hat{e}_{\rm ph})}\right]^{1/2}.
 \ee

 \subsection{Optical depth due to the companion starlight }

Starlight photons coming from the companion mid-A supergiant at a
temperature $T\approx 8500 \ {\rm K}$ \cite{Hillwig,Barnes} can
cause absorption of gamma-rays. To see this clearly let us consider
some simple geometry.

In Fig. \ref{Figrender} the observer is assumed to lay in the
$xz$-plane and the gamma-ray path is described by the vector
 \be
{\vec \rho_\gamma}=\rho_\gamma(\hat{x}\sin i + \hat{z}\cos i)=
{\rho}_\gamma \hat{e}_\gamma ,\label{gammaunit}
 \ee
 with $i=78^\circ$ as mentioned.
 The position of the star is given by
 $\vec{a}= a(\hat{x} \cos \phi  +  \hat{y} \sin
 \phi)$ and we suppose that the gamma-ray is produced
 in the jet at
 \be
\vec{z}_{\rm j}= z_{\rm j}(\hat{x}\sin \theta \cos \psi  +
\hat{y}\sin \theta \sin \psi + \hat{z}\cos \theta ),
 \ee
where $\psi$ is the precessional phase. The position where the
interaction with the soft photon takes place is indicated by
 \be
\vec{r}_\gamma= \vec{z}_{\rm j}+ \vec{\rho}_\gamma. \label{gammapos}
 \ee

 In this case, we set $\vec{l}_{\rm ph}=(\vec{r}_\gamma-\vec{a})$
as the vector connecting the center of the star with the gamma-ray,
and we call $\alpha$ the angle that it makes with $\hat{e}_\gamma$.

At each particular gamma-ray position, a convenient reference system
can be chosen to perform the integration in the angles $\theta'$ and
$\phi'$ in expression (\ref{dtau}).  This is done, as in Ref.
\cite{DUB05}, orientating the $z'$-axis along the vector
$\vec{l}_{\rm ph}$. Accordingly, the angle $\phi'$ varies between
$\phi'_{\rm min}=0$ and $\phi'_{\rm max}=2\pi$ and $\theta'$ varies
between $\theta'_{\rm min}= 0$ and $\theta'_{\rm max}=
\arccos(\sqrt{1-(R_\star/l_{\rm ph})^2})$. Further orientating the
$x'$-axis so that the gamma-ray path is contained in the
$x'z'$-plane, the gamma-ray direction can be written as
$\hat{e}_\gamma=(\sin \alpha,0,\cos\alpha)$ and hence,
$\hat{e}_\gamma\cdot\hat{e}_{\rm ph}=
\sin\alpha\sin\theta'\cos\phi'+ \cos\alpha\cos\theta'$. However, the
dot-product involved for the evaluation of the cross section
(\ref{sigee}) and the minimum photon energy (\ref{Emin}) is
approximated by $\cos\alpha$, which does not depend on the primed
angles. This last simplification that permits the analytic
integration on $\phi'$ and $\theta'$, means only that in order to
compute the cross section and the minimum energy, the starlight
photons are all considered as being originated at the center of the
star.


Integration in $\theta'$ and $\phi'$ of expression (\ref{dtau}) thus
gives
 \be
\frac{d\tau_{\star}}{dE d\rho_\gamma}= 2\pi \
n_\star(E)\ \sigma_{\gamma\gamma}(E,E_\gamma)  \nonumber \\
\times\left[\cos \theta'_{\rm min}- \cos \theta'_{\rm max} \right.\label{dtaustar}\\
\left. -\frac{\cos\alpha}{2} (\sin^2 \theta'_{\rm max}-\sin^2
\theta'_{\rm min}) \right]. \nonumber
 \ee
The density of radiation from the star can be approximated as usual
by
 \be
n_\star(E)= \frac{2E^2}{(h \ c)^3( e^{E/kT_\star}- 1)} (\rm ph \
cm^{-3} {\rm erg}^{-1}{\rm sr}^{-1}). \label{nbb}
 \ee
Further integrating in target photon energy $E$ and along the
gamma-ray path $\rho_\gamma$ yields the resulting starlight
contribution to the optical depth
 \be
 \tau_\star= \int_{E_{\rm min}}^\infty dE
 \int_0^\infty d\rho_\gamma \frac{d\tau_{\star}}{dEd\rho_\gamma}
 \label{taustar},
 \ee
where
 \be
 E_{\rm min}= \frac{2 (m_e c^2)^2}{E_\gamma(1- \hat{e}_\gamma\cdot\hat{e}_{\rm
 ph})}.\label{Emin}
 \ee

\begin{figure}[h]
\includegraphics[trim = 10mm 6mm 0mm 8mm, clip, width=9cm,angle=0]{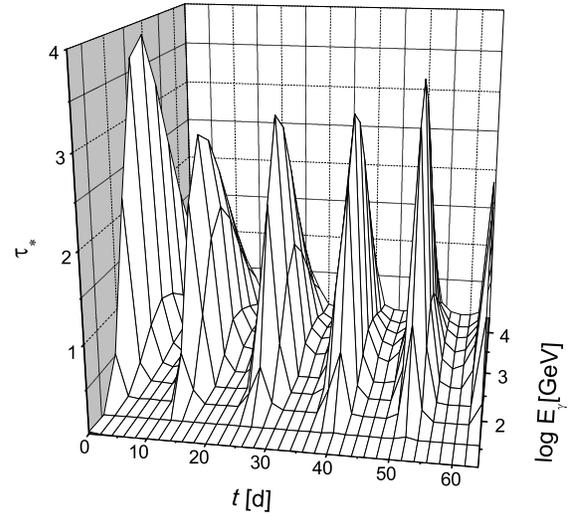}
\caption{Starlight optical depth.} \label{Figtaustara}
\end{figure}

Once one relates the position vectors with time, one can obtain
$\tau_\star(E_\gamma, t)$ as shown in Fig.\ref{Figtaustara}. These
gamma-rays were assumed to be originated mostly near an injection
point at the base of the approaching jet, $z_0= R_0/\tan{\xi}\simeq
1.3 \times 10^{9}$cm, where $\xi\simeq 0.6$ is the half opening
angle of the jet (see Ref. \cite{marshall}), and the initial jet
radius is taken to be $R_0= 10 GM_{\rm bh}/c^2$. It can be seen from
Fig.\ref{Figtaustara} that this component of the opacity is
notoriously more important when the star is in the foreground of the
emission point, namely when it crosses the $x$-axis.


We note that to perform the intgral in Eq.(\ref{taustar}), we have
assumed for simplicity that the radiation density given by
Eq.(\ref{nbb}) was artificially valid inside the star. However, as
will be shown below, the absorption caused by $\gamma N$
interactions is by far the dominating one if the gamma-ray path
directed to Earth has to travel through the star. This implies that,
although the peaks in Fig. \ref{Figtaustara} could look different,
an accurate expression for the radiation density inside the star
will not affect the total optical depth.

\subsection{Optical depth due to the emission from the extended disk} \label{secUV}

The extended disk wind is believed to be the origin of both the
observed mid-IR and UV emission. In the first case, the reported
free-free emission was detected in the range of wavelengths $2$-$12
\ \mu {\rm m}$ for which we adopt the following flux density fit
\cite{Fuchs05},
 \be
 F_{\rm ph}= 2.3\times 10^{-23}
\left(\frac{\lambda}{\mu \rm m}\right)^{-0.6} ({\rm erg \ s^{-1}
cm^{-2} Hz^{-1}}),
 \ee
which corresponds an emitting region of radius $R_{\rm out}=50
R_\odot$. We need to estimate the radiation density of these mid-IR
photons, i.e., the number of photons per unit volume, per unit
energy per unit solid angle spanned by the emitting region. The
corresponding mid-IR photon flux per unit frequency arriving at
Earth can be written as
 \be
\frac{dN_{\rm ph}}{dt \ dA  \ d\nu }=\frac{F_{\rm ph}}{E} ({\rm ph \
cm}^{-2}{\rm s}^{-1}{\rm Hz}^{-1}).
 \ee
Considering that $dE/d\nu= h$, $dl_{\rm ph}/dt=c$, and that $dl_{\rm
ph}dA=dV$, the radiation density coming from the whole emitting
region is
 \be
 \frac{dN_{\rm ph}}{dV dE}= \frac{F_{\rm ph}}{E \ hc} ({\rm ph \ cm}^{-3}{\rm
 erg}^{-1}).
 \ee
Since the adopted fit for $F_{\rm ph}$ corresponds to the particular
precessional phase $\psi_{\rm IR}= 0.31\times 2\pi$ (see Ref.
\cite{Fuchs05}), in that case, our line of sight makes an angle
$i_{\rm j,IR}$ with the normal to the disk midplane such that $\cos
i_{\rm j,IR}= \sin i \sin \theta \cos\psi_{\rm IR}+ \cos i \cos
\theta$. Then, the area of the emitting region $\pi R_{\rm out}^2$
spans a solid angle $\Delta\Omega= (\pi R_{\rm out}^2\cos i_{\rm
j,IR})/d^2$ as viewed from Earth, and we can estimate the
corresponding radiation density as
 \be
n_{\rm IR}& \approx & \frac{1}{\Delta\Omega}\left(\frac{dN_{\rm
ph}}{dV
dE}\right)\\
&=& \frac{F_{\rm ph}}{E \ hc \ \pi \cos i_{\rm j,IR}}
\left(\frac{d}{R_{\rm out}}\right)^2 ({\rm ph \ cm}^{-3}{\rm
erg}^{-1}{\rm sr}^{-1}).
 \ee



\begin{figure}[h]
\includegraphics[trim = 10mm 6mm 0mm 8mm, clip, width=9cm,angle=0]{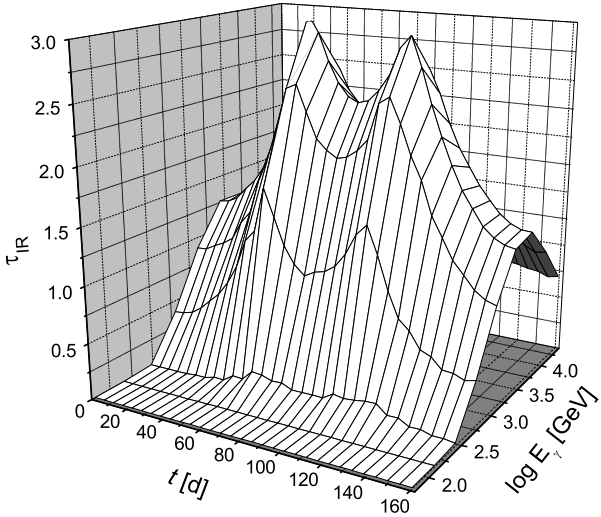}
\caption{Mid-IR optical depth as a function of time and gamma-ray
energy for a gamma-ray produced at a height $z_0$ on the approaching
jet. }\label{FigtauIRa21}
\end{figure}

We proceed to integrate the mid-IR contribution of the optical depth
$\tau_{\rm IR}$ using the method described in Appendix A for a
disk-like emitting zone. The inner radius of this zone is taken as
the inner radius of the extended disk, $R_{\rm in}\approx 2GM_{\rm
bh}/v_w^2$, as suggested in Ref. \cite{BKP06}. The obtained result
is shown in Fig. \ref{FigtauIRa21} for gamma-rays originated at the
base of the approaching jet.

As for the UV emission, it was detected in the range of wavelengths
$\sim (1000- 10000){\AA}$ \cite{Gies02} and, as a first approach, we
consider an associated blackbody temperature $T_{\rm UV}= 21000 \
\rm{K}$ for the disk in an edge-on state, with a corresponding
radius of the emitting zone $R_{\rm out}= 33 R_\odot$. Therefore, we
consider a radiation density for the UV emission as the one given in
expression (\ref{nbb}) with $T_\star\rightarrow T_{\rm UV}$.

The integration of this contribution to the total optical depth is
also performed as it is described in Appendix A. The obtained
contribution is shown in Fig. \ref{FigtauUVa21} as a function of
time and gamma-ray energy.

It can be noted from this plot and from Fig. \ref{FigtauIRa21} that
the maximum absorption occurs near the two times $t\sim 55$ d and
$t\sim 110$ d. Since the mid-IR and UV emissions are treated in this
first analysis as being originated in the midplane of the extended
disk, these times correspond to the cases when the gamma-rays
originated at $z_0$ and directed to Earth run almost parallel to the
mentioned emitting zone. This implies that $\gamma\gamma$ collisions
will occur at larger angles resulting in higher cross sections and
smaller threshold energies along a significant part of the gamma-ray
path. At intermediate times, $55{\rm \ d}<t<110{\rm \ d}$, the
mentioned gamma-rays have to pass through the midplane of the
extended disk, and once they have done so, the collision angles are
smaller, the threshold energy is larger, and hence the absorption is
lower.

As for the dependence with the gamma-ray energy, it can be seen from
Figs. \ref{FigtauIRa21} and \ref{FigtauUVa21} that the UV optical
depth becomes important for $E_\gamma \gtrsim 50$ GeV, while the
mid-IR contribution will be significant for $E_\gamma \gtrsim 500$
GeV.


\begin{figure}[h]
\includegraphics[trim = 10mm 6mm 0mm 8mm, clip, width=9cm,angle=0]{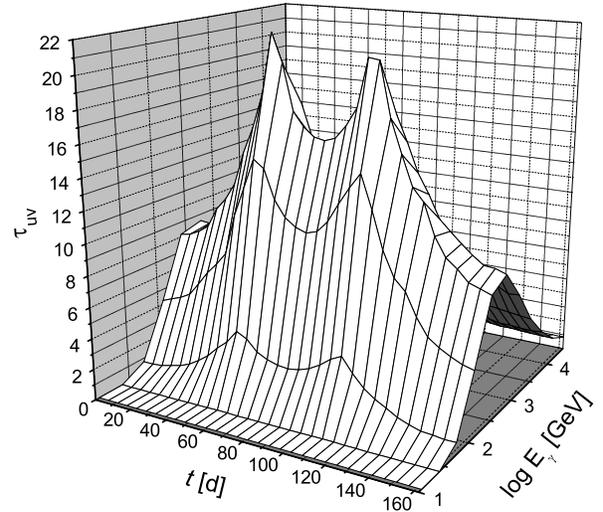}
\caption{UV optical depth as a function of time and gamma-ray energy
for a gamma-ray produced at a height $z_0$ on the approaching jet.
}\label{FigtauUVa21}
\end{figure}

\subsection{Optical depth due to $\gamma N$ interactions.}

The nucleons of the star and the thick extended disk can also absorb
gamma-rays and photo-produce pions. We consider the cross section as
in Ref. \cite{AD03},
 \be
\sigma_{\gamma N}= \left\lbrace\begin{array}{cc}
  340 \ \mu{\rm b}& {\rm \ for \ } 200\;{\rm MeV}<E_\gamma <500\;{\rm MeV} \\
  120 \ \mu{\rm b}& {\rm \ for \ } E_\gamma \geq 500\;{\rm MeV}
\end{array}\right.,
 \ee
where the first case corresponds to the single pion channel and the
second case to the multi-pion channel.

 Considering the unit vector $\hat{Z}=(\sin \theta \cos \psi,\sin
\theta \sin \psi,\cos \theta)$, which is perpendicular to the
midplane of the disk wind, and the angle it makes with the gamma-ray
position, $\theta_Z=
\arccos({\hat{Z}\cdot\vec{r}_\gamma/r_\gamma})$, the density of the
extended disk wind at the point $\vec{r}_\gamma$ is estimated to be
 \be
\rho_{\rm w}(r_\gamma,\theta_Z)= \frac{\dot{M}_{\rm w}}{v_{\rm w}
\Delta\Omega r_\gamma^2}
\Theta\left[\theta_Z-\left(\frac{\pi}{2}-\alpha_{\rm w}\right)
\right]\Theta\left[\left(\frac{\pi}{2}+\alpha_{\rm w}\right)
-\theta_Z\right]. \label{nNdisk}
 \ee
Here, $\Theta$ is the Heaviside step function and the solid angle
element is related to the disk wind half opening angle $\alpha_{\rm
w}$ by $\Delta\Omega= 4\pi\sin\alpha_{\rm w}$.

As for the star, we suppose that it has a density given by
  \be
 \rho_\star(r)= \frac{M_\star}{4\pi R_\star r^2} \Theta(r-R_\star),
  \ee
where $r$ represents the distance to the center of the star.


The optical depth due to $\gamma N$ interactions with the mentioned
nucleons can be estimated as
 \be
\tau_{\gamma N}(\vec{z}_{\rm j})= \int_0^\infty d\rho_\gamma
\sigma_{\gamma N} \frac{\left(\rho_\star+ \rho_{\rm w}\right)}{m_p}.
\label{taugammaN}
 \ee

\begin{figure}[h]
\includegraphics[trim = 7mm 5mm 0mm 8mm, clip, width=9cm,angle=0]{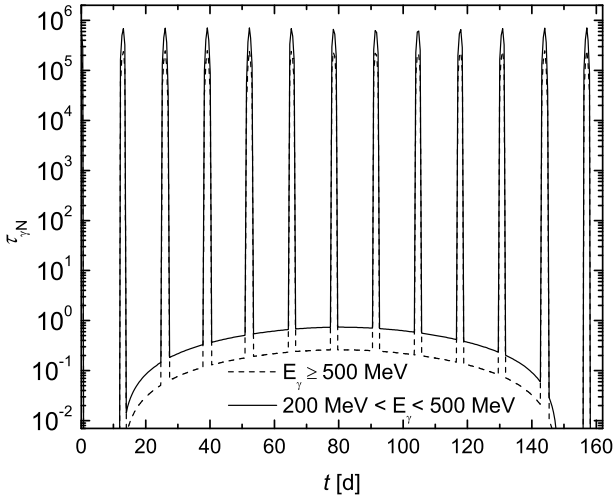}
\caption{$\gamma N$ optical depth as a function of time for $200 \
{\rm MeV}<E_\gamma<500$ MeV (solid line) and $E_\gamma> 500$ MeV
(dashed line). The sharp peaks are due to the orbital modulation.
}\label{FigtauGNaStar}
\end{figure}

The obtained result is shown in Fig. \ref{FigtauGNaStar} as a
function of time for $200 \ {\rm MeV}<E_\gamma<500$ MeV and
$E_\gamma>500$ MeV. As it can be seen from this plot, periodic peaks
of high absorption appear when the star is in the foreground. We
note that the absorption caused by the disk wind is rather low. In
particular, it vanishes when the approaching jet is most open to us
(for $t\lesssim 12$ d and $t\gtrsim 148$ d ) because in those cases
the gamma-ray path originated at the base of the jet never travels
through the disk wind.

 \subsection{Total optical depth}

Analyzing the different contributions, it can be noted that below
$E_\gamma= 20$ GeV, the only relevant source of absorption is due to
the $\gamma N$ interactions with the nucleons of the extended disk
and the star.

\begin{figure}[h]
\includegraphics[trim = 5mm 5mm 9mm 8mm, clip, width=9cm,angle=0]{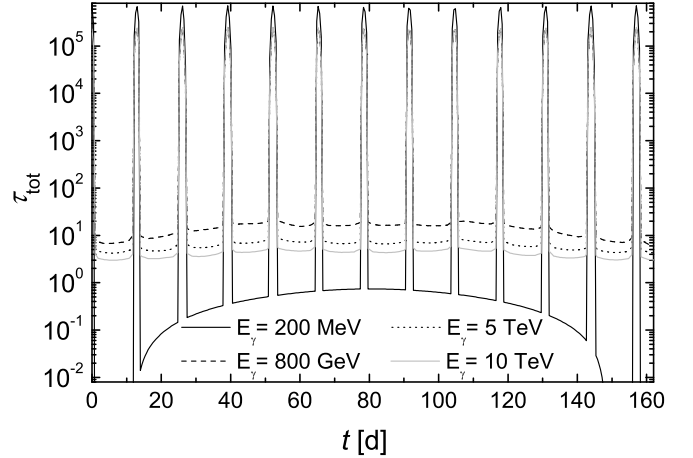}
\caption{Total optical depth for gamma-rays with as a function of
time for $E_\gamma= 200$ MeV (solid line), $E_\gamma= 800$ GeV
(dashed line), $E_\gamma= 5$ TeV (dotted line), and $E_\gamma= 10$
TeV (gray solid line). }\label{Figtautota21}
\end{figure}

If we add up all the contributions calculated above for the case of
gamma-rays originated at a height $z_0$ in the approaching jet, we
can obtain the total optical depth which is shown in Fig.
\ref{Figtautota21} as a function of time and gamma-ray energy. It
can be seen clearly from this plot that the absorption caused by
$\gamma N$ interactions with the nucleons of the star is important
along all the range of gamma-ray energies studied ($200$ MeV - $20$
TeV).

 \subsection{Dependence of the absorption on the gamma-ray
injection point}

 We shall now consider the cases where most of the gamma-rays
are produced at a given height $z_{\rm j}>z_0$ along the approaching
jet. The calculation of the various components of the optical depth
can be performed as described in the previous section substituting
$z_0$ by $z_{\rm j}$.

\begin{figure}[h]
\includegraphics[trim = 11mm 1mm 8mm 2mm, clip, width=9cm,angle=0]{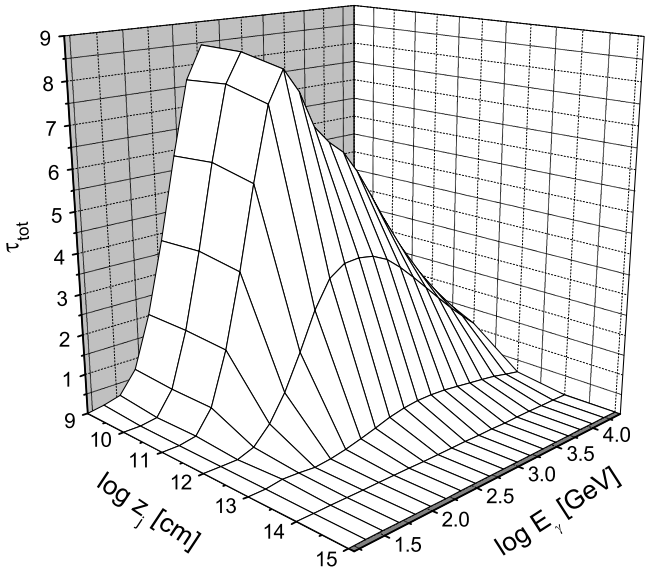}
\caption{Total optical depth at $t=162$ d for gamma-rays originated
with energy $E_\gamma$ at a distance $z_{\rm j}$ from the compact
object along the approaching jet. }\label{Figtau1dA}
\end{figure}

\begin{figure}[h]
\includegraphics[trim = 11mm 5mm 8mm 8mm, clip, width=9cm,angle=0]{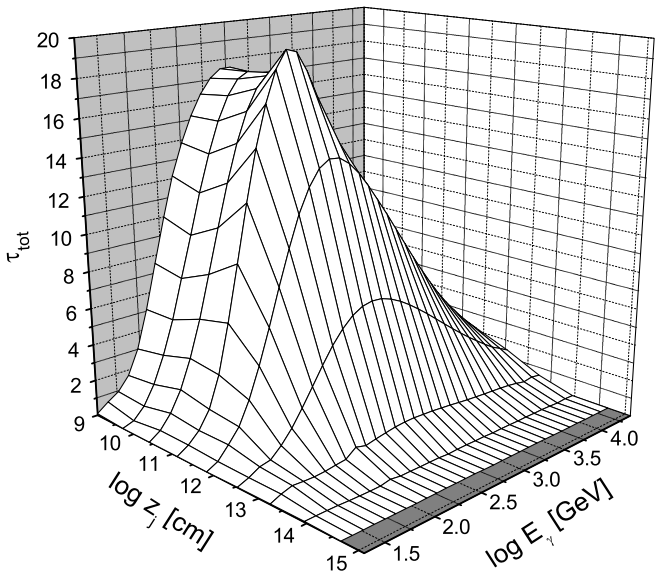}
\caption{Total optical depth at $t=81$ d for gamma-rays originated
with energy $E_\gamma$ at a distance $z_{\rm j}$ from the compact
object along the approaching jet. }\label{Figtau80dA}
\end{figure}

We show the obtained total optical depth as a function of $z_{\rm
j}$ and the gamma-ray energy in Figs. \ref{Figtau1dA} and
\ref{Figtau80dA} for $t\sim 162$ d and $t\sim 81$ d respectively,
representing the cases when the absorption is maximum and minimum.
Since at these times the star is not in the foreground, the optical
depth is basically the one corresponding to the UV and mid-IR
emission from the disk wind. At $t=162$ d, when the approaching jet
is mostly open to our direction, the optical depth decreases with
$z_{\rm j}$ because as $z_{\rm j}$ increases, the distance to be
traveled by the gamma-ray directed to Earth over which
$\gamma\gamma$ interactions are important decreases.

At $t=81$ d there is an initial rise of the optical depth with
$z_{\rm j}$. This is because at this particular time, the
\textit{approaching jet} as was referred here for simplicity, is
actually not approaching the Earth but directed away from it. Then,
as $z_{\rm j}$ increases, the distance along which $\gamma\gamma$
collisions are relevant increases for the gamma-rays coming in our
direction. However, for still higher values, $z_{\rm j}\gtrsim
10^{11}$cm, the optical depth decreases with $z_{\rm j}$ since the
density of UV radiation from the extended disk wind is lower.

\section{SIGNATURES ON A GAMMA-RAY SIGNAL}

In the present work we have not specified any particular gamma-ray
emission process. Regardless of this, one possibility is that since
the particle densities are expected to be highest in the inner jet,
most of the gamma-rays could be produced at a short distance
$z_{\rm j}$ from the compact object. In view of the recent detection
of gamma-rays from similar systems, such as LS I +61 303, LS 5039,
and Cyg X-1 \cite{LSI,aharonian,CYG}, and
in an attempt to give an idea of the effect that the obtained
optical depth may cause on an out-coming gamma-ray flux, we can
assume that most of the radiation is produced near the base of the
jets (at $z_{\rm j}\sim z_0$). In these conditions, if the spectrum
of produced gamma-rays follows a power law as
 \be
 J_\gamma = K_\gamma E_\gamma^{-2} ({\rm ph \ erg}^{-1} {\rm sr}^{-1}{\rm cm}^{-2}{\rm
 s}^{-1}),
 \ee
where $K_\gamma$ is a constant, the total luminosity within an energy range $(E_\gamma^{\rm min},E_\gamma^{\rm max})$ is
\be
 L_\gamma= \Delta A \int_{E_\gamma^{\rm min}}^{E_\gamma^{\rm max}}
 dE_\gamma E_\gamma J_\gamma(E_\gamma),\label{lumgamma}
 \ee
where $\Delta A$ is the element of area of the emitting region near
the injection point. It follows, then, that
 \be
 K_\gamma \Delta A= \frac{L_\gamma}{\ln{\frac{E_\gamma^{\rm max}}{E_\gamma^{\rm
 min}}}},
 \ee
which allows us to estimate the corresponding photon flux to be
detected at the Earth as
 \be
 \Phi_\gamma(t)= \frac{\Delta A}{4\pi d^2} \int_{E_\gamma^{\rm min}}^{E_\gamma^{\rm max}}
 dE_\gamma  J_\gamma(E_\gamma) e^{-\tau_{\rm tot}(t,E_\gamma)}.
 \label{gammaflux}
 \ee

\begin{figure}[h]
\includegraphics[trim = 0mm 0mm 0mm 8mm, clip, width=9cm,angle=0]{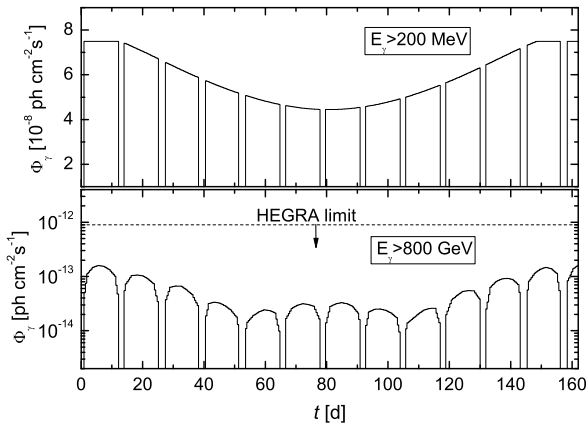}
\caption{Gamma-ray flux a function of time for $E_\gamma>200$ MeV
(upper panel) and for $E_\gamma>800$ GeV (lower panel) as compared
to the upper limit given by HEGRA. The assumed gamma-ray luminosity
emitted is $L_\gamma= 10^{36} {\rm erg \ s}^{-1}$ and the maximum
energy is $E_\gamma^{\rm max}=20$ TeV } \label{FigGfluxcrude}
\end{figure}

For illustration, we assume that the equivalent isotropic gamma-ray
luminosity between $E_\gamma^{\rm min}= 200$ MeV and $E_\gamma^{\rm
max}= 20$ TeV is $L_\gamma\approx 10^{36} \rm{erg \ s}^{-1}$. The
resulting flux is shown in the upper panel of Fig.
\ref{FigGfluxcrude}, whereas the flux corresponding to energies
$E_\gamma> 800$ GeV is shown in the lower panel as compared to the
upper limit given by HEGRA \cite{HEGRA}. In the first case the
obtained mean photon flux is
 \be
\left. \langle \Phi_\gamma\rangle\right|_{E_\gamma>200 {\rm MeV}}= 5
\times 10^{-8} \ {\rm ph \ cm}^{-2}{\rm s}^{-1},
 \ee
and in the second case,
 \be
\left. \langle \Phi_\gamma\rangle\right|_{E_\gamma>800 {\rm GeV}}=
4.7 \times 10^{-14} \  {\rm ph \ cm}^{-2}{\rm s}^{-1},
 \ee
which is below the HEGRA cut, $\Phi_\gamma^{\rm lim}= 8.9 \times
10^{-13} {\rm \ ph \ cm}^{-2}{\rm s}^{-1}$.

 We emphasize that these values are assumed only to make a
qualitative description of a possible gamma-ray signal and that
their true values may be obtained in a more detailed study regarding
the emission process that could operate.


\section{DISCUSSION}

As it is reasonable to expect, the most noticeable absorption
signature that can be imprinted on a gamma-ray flux from the inner
regions of SS433 is given by the regular and dramatic absorption
caused by the star when it crosses the line of sight. This effect
takes place approximately once every $13$ days during the time that
the star blocks the emitting region ($\approx 2$ d). In practice,
this could serve to determine the size of the companion star as well
as the mass distribution of the stellar atmosphere as long as
sufficient time resolution can be achieved in the $\gamma$-ray
lightcurves. It seems possible that, given the little absorption
that corresponds to energies below $\sim 50$ GeV and the fact that
more events are expected for lower energies, the GLAST instrument
could detect a signal possibly with the time behavior predicted
here. Another detectable absorption feature to observe at these
energies corresponds to the $\gamma N$ interactions with the
nucleons of the extended disk. This is expected to cause, because of
the precession, a long term modulation of the signal reducing the
maximum flux by a significant fraction as shown in the upper panel
of Fig. \ref{FigGfluxcrude}, provided that the bulk of the emission
is produced near $z_0$.

As for higher energies, absorption through $\gamma\gamma$
interactions also becomes important. The dominant contribution is
caused by the UV emission from the extended disk for $E_\gamma
\gtrsim 50$ GeV, while the mid-IR photons there originated can cause
rather significant attenuation for $E_\gamma \gtrsim 500$ GeV. The
absorption due to the starlight photons is significant when the star
is nearly in the foreground, almost eclipsing the black hole. This
can be seen in the lower panel of Fig. \ref{FigGfluxcrude}, where
the gamma-ray signal decreases and increases quite smoothly before
and after the star has actually eclipsed the emission zone.

The fact that HEGRA has failed to detect anything greater than
$\Phi_\gamma^{\rm lim}=8.9 \times 10^{-13}{\rm ph \ s}^{-1}{\rm
cm}^{-2}$ for $E_\gamma>800$ GeV, implies that maybe SS433 is not
intrinsically luminous at these energies, or that gamma-rays could
be actually produced at an inner injection point as illustrated here
with a given luminosity, but nothing could be detected at the Earth
because of absorption. If the assumed luminosity in gamma-rays is
similar to the real intrinsic luminosity, a detection at these
energies seems more difficult but not impossible with the
forthcoming Cherenkov telescopes MAGIC II and VERITAS.


Finally we remark again that the conclusions drawn in the present
study are independent of the gamma-ray producing mechanism. Taking
into account reprocessing of gamma-rays due to cascading effects
\cite{wlodek}, would not either alter the absorption patterns here
predicted. In case of an hadronic mechanism for gamma-ray production
in the source (e.g. \cite{rom03}), neutrino emission should be
expected, as pointed out by several authors \cite{hugo2,aha06}.
Contrary to gamma-rays, the neutrino signal is not affected by
absorption, and its detection by instruments like IceCube might
yield light on the relativistic particle content of the inner jets.

A comprehensive study, including the description of the physical
mechanisms allowing the high-energy emission processes in SS433 will be
presented elsewhere.\\

\textit{Acknowledgements} We thank V. Bosch-Ramon for useful
comments and an anonymous referee for constructive criticism. G.E.R.
is supported by the Argentine agencies CONICET (PIP 5375) and ANPCyT
(PICT 03- 13291 BID 1728/OC-AR). G.E.R. also acknowledges support
from the Ministerio de Eduaci\'on y Ciencia (Spain) under grant
AYA2007-68034-C03-01, FEDER funds. H.R.C. is supported by CNPq and
FUNCAP, Brazil, and M.M.R. is supported by CONICET, Argentina.
M.M.R. is also grateful to O. A. Sampayo for useful discussions
relevant to the present work.

\appendix \section{Calculation of the gamma-ray absorption near a
disk-like emitting region}

We consider the absorption of gamma-rays as they travel near a disk
that emits soft photons. This disk, with an inner radius $R_{\rm
in}$ and an outer radius $R_{\rm out}$, lies in a plane
perpendicular to the jets, where the gamma-rays emerge. For
gamma-rays produced with energy $E_\gamma$, traveling in a direction
$\hat{e}_\gamma$, the differential optical depth due to photons with
a radiation density $n_{\rm ph}$ and a direction given by
$\hat{e}_{\rm ph}$, can be written as (e.g \cite{BK95})
 \be
d \tau_{\gamma\gamma}= ( 1- \hat{e}_{\rm ph}\cdot\hat{e}_{\gamma} )
n_{\rm ph} \sigma_{\gamma\gamma} \frac{\cos \eta R_{\rm d} dR_{\rm
d} d\phi_{\rm d} }{l_{\rm ph}^2} d\rho_\gamma dE \label{dtaudisk}.
 \ee
Here the $R_{\rm d}$ is the length of the vector $\vec{R}_{\rm d}$
signaling a point in the disk, $\phi_{\rm d}$ is its corresponding
azimuthal angle, and $l_{\rm ph}$ is the magnitude of the vector
$\vec{l}_{\rm ph}$ connecting the point on the disk to the gamma-ray
position. This vector makes an angle $\eta$ with the normal of the
disk plane. All the angles and vectors involved in (\ref{dtaudisk})
are then to be expressed in a new coordinate system with its
$Z$-axis oriented along the axis of the approaching jet.

It is worth noting that the radiation density $n_{\rm ph}$ appearing
in expression (\ref{dtaudisk}) represents the number of soft photons
per unit volume, per unit energy, per unit solid angle. Hence, for
photons originated on a differential area element $da=R_{\rm
d}dR_{\rm d} d\phi_{\rm d}$ on the disk plane, the differential
solid angle spanned is $d\Omega= (da \cos{\eta})/l_{\rm ph}^2$,
which appears in expression (\ref{dtaudisk}).

In the system fixed to the compact object the unit vector
$\hat{e}_\gamma$ is given by Eq. (\ref{gammaunit}) in terms of the
unit vectors $\hat{x}$, $\hat{y}$ and $\hat{z}$, and the basis of
the new system can be obtained by transforming them according to
 \be
 R= \left(
      \begin{array}{ccc}
        \cos \psi \cos \theta & -\sin \psi  & \cos \psi \sin \theta \\
        \sin \psi \cos\theta & \cos\psi & \sin\psi \sin\theta \\
        -\sin\theta & 0 & \cos\theta \\
      \end{array}.
    \right)
 \ee
Therefore, the unit vectors of the new coordinate system can be
expressed in terms of fixed basis as
 \be
 \hat{X}&=& \hat{x} \cos\theta \cos\psi   +\hat{y}  \cos\theta \sin\psi
- \hat{z} \sin\theta \\
 \hat{Y}&=& -\hat{x} \sin\psi   +\hat{y}  \cos\psi
 \\
 \hat{Z}&=& \hat{x} \sin\theta \cos\psi   +\hat{y}  \sin\theta \sin\psi
 + \hat{z}\cos\theta,
 \ee
and the unit vector $\hat{e}_\gamma$ can be written as
 \be
\hat{e}_\gamma&=& \hat{X}\left( \hat{e}_\gamma \cdot\hat{X}\right) +
\hat{Y}\left( \hat{e}_\gamma\cdot\hat{Y}\right) + \hat{Z}\left(
\hat{e}_\gamma\cdot\hat{Z}\right) \nonumber \\
\hat{e}_\gamma&=& \hat{X}\left( \cos\theta \sin\psi \sin i - \sin
\theta \cos i \right)\nonumber\\ &-& \hat{Y}\left( \sin\psi \sin
i\right)  \\ &+& \hat{Z}\left( \sin\theta \cos\psi \sin i+ \cos
\theta \cos i \right). \nonumber
 \ee

Since the position of the gamma-ray is
 \be
\vec{r}_\gamma= \hat{e}_\gamma\rho_\gamma  + \hat{Z}z_{\rm j}
\nonumber
 \ee
 and the position on the disk is
  \be
   \vec{R}_{\rm d}= R_{\rm d}\left(\hat{X}\cos \phi_{\rm d} + \hat{Y}\sin \phi_{\rm d}
   \right),
  \ee
the vector $\vec{l}_{\rm ph}$ can be obtained as $\vec{l}_{\rm ph}=
\vec{r}_\gamma- \vec{R}_{\rm d}$.

The resulting optical depth is given by the quadruple integral
 \be
\tau_{\gamma\gamma}= \int_0^\infty d\rho_\gamma \int_0^{2\pi}
d\phi_{\rm d} \int_{R_{\rm in}}^{R_{\rm out}} dR_{\rm d}
\int_{E_{\rm min}}^{E_{\rm max}} dE \frac{d \tau_{\gamma\gamma}}{d
\rho_\gamma d \phi_{\rm d} d R_{\rm d} d E }\label{quadru},
 \ee
 where
 \be
 E_{\rm min}= \frac{2 (m_e c^2)^2}{E_\gamma(1- \hat{e}_\gamma\cdot\hat{e}_{\rm
 ph})}.
 \ee

The integral can be performed using a Monte Carlo method, that is,
introducing the variables $x_\rho, \ x_\phi, \ x_R$, and $x_E$ as
 \be
 \rho_\gamma&=& \rho_{\rm 1} x_\rho\\
 \phi_{\rm d}&=& 2\pi x_\phi\\
 R_{\rm d}&=& R_{\rm in}+ (R_{\rm out}-R_{\rm in}) x_R\\
 E&=& E_{\rm min}+ (E_{\rm max}-E_{\rm min}) x_E,
 \ee
then the integral (\ref{quadru}) can be written as
 \be
\tau_{\gamma\gamma}= \int_0^1 d x_\rho \int_0^1 d x_\phi \int_0^1 d
x_R \int_0^1 d x_E f(x_\rho,x_\phi,x_R,x_E),\label{quadru2}
 \ee
where
 \be
f(x_\rho,x_\phi,x_R,x_E)=  2\pi  \rho_1 (R_{\rm out}-R_{\rm in})
\times \nonumber \\ (E_{\rm max}-E_{\rm min})  \frac{d
\tau_{\gamma\gamma}}{d \rho_\gamma d \phi_{\rm d} d R_{\rm d} d E }.
 \ee
Here the upper limit of the integration in $\rho_\gamma$ is taken as
$\rho_1\approx 10 R_{\rm out}$ where the integrand is significative.

 The method relies on the fact that, by the Mean-Value Theorem,
the integral can be approximated by
 \be
\tau_{\gamma\gamma}\approx \frac{1}{N}\sum_{i=1}^N
f\left(x_\rho(i),x_\phi(i),x_R(i),x_E(i)\right),
 \ee
where each variable takes a random number between $0$ and $1$. The
right hand side is then the statistical average of $N$ evaluations
of the function $f$. The error goes like $1/\sqrt{N}$, which in this
case of four integration variables makes it a quite accurate method
as compared to the iterative ones (see e.g. \cite{CollPhys}).

\end{document}